\documentclass[12pt
 ]{article}
\usepackage{graphicx}

\begin{document}
\begin{titlepage}
\centerline{\Large {\bf  Liouville-von Neumann approach and
time-dependent}} \vspace{1.5mm} \centerline{\Large {\bf  Gaussian
approximation }} \vspace{1.5mm}
\vspace{10mm}
\centerline{ Hyeong-Chan Kim$^\dagger$
\footnote{E-mail: \texttt{hckim@phya.yonsei.ac.kr}} and Jae Hyung
Yee$^{\dagger\star}$ \footnote{E-mail:
\texttt{jhyee@phya.yonsei.ac.kr}}}

\centerline{$^\dagger$\it Institute of Basic Science, Yonsei
University, Seoul 120-749, Korea}
\centerline{$^\star$\it Institute of Physics and Applied Physics,
Yonsei University, Seoul 120-749, Korea}
\vskip 5mm \centerline{(\today)}
\setlength{\footnotesep}{0.5\footnotesep}

\vskip 10mm

\begin{abstract}
We show that Liouville-von Neumann approach to quantum mechanical
systems, which demands the existence of invariant operators,
reproduces the time-dependent variational Gaussian approximation.
We find the effective action of the time-dependent systems and
show that many aspects of the dynamics are independent of the
details of time evolution, e.g., the squeezing of the
wave-function is determined by the effective potential of the
final stage of time-evolution.
\end{abstract}

\vspace{5mm}

\indent\indent\hspace{4mm}  {\bf Keywords:} Liouville-von Neumann
approach, Gaussian, slow-rolling \vspace{3mm}
\indent\indent\hspace{4mm}


\end{titlepage}

\setcounter{footnote}{0}

\section{Introduction}

The interests in searching for a method which effectively
describes the real-time quantum dynamics has been continually
increased, in relation with the quantum theory out of thermal
equilibrium~\cite{jackiw}, the inflationary scenario of the
Universe~\cite{guth}, the dynamic process in phase
transitions~\cite{zurek}, the formation of topological defects in
variety of phenomena in condensed matter physics~\cite{bary}.
Understanding quantum mechanical evolution of inflaton field in
the new inflationary universe is vital for correct determination
of the inflationary epoch's period~\cite{guth2}.

A time-dependent variational approximation for quantum dynamics
was developed in Schr\"{o}dinger picture~\cite{cooper,Pi1}. The
Liouville-von Neumann (LvN) approach of the time-dependent
harmonic~\cite{lvn} and driven harmonic~\cite{kim} oscillators
have been developed to obtain the exact time-dependence of quantum
states, and the methods are generalized to the cases of anharmonic
oscillator, scalar inflaton field dynamics~\cite{spkim}, and
driven anharmonic oscillator~\cite{kim2}. LvN approach has a merit
that it can be directly applied to both time-independent and
time-dependent quantum systems without any modification,  because
it is based on the operator level equations.

The purpose of this paper is to extend the approach developed in
Ref.~\cite{spkim} to include arbitrary potential, so that one can
develop a general Gaussian approximation using the LvN method.
This approach has two merits: 1) it can describe the
time-evolution of systems from the start to the end. 2) In LvN
approach, it is easy to define the general Fock space because it
directly identifies the creation and annihilation operators.

The organization of this paper is as follows. In Sec. II, the LvN
approach is elaborated to be applicable to the case of general
potential $V(\hat{q},t)$ in obtaining the Gaussian approximation,
and is applied to find the Fock space and the time-dependent
expectation values of some operators for $V(\hat{q},t)$. Sec. III
is devoted to the study of a model with slow rolling potential. In
Sec. IV, we summarize our results and present some discussions.

\section{Gaussian approximation in Liouville-von Neumann approach}

The quantization of a time-dependent anharmonic oscillator using
the LvN approach was studied in Ref.~\cite{spkim}. It was shown
that their approach for symmetric potential is equivalent to the
variational Gaussian approximation and the mean-field,
Hartree-Fock method. In this section, we develop LvN method for a
general potential and show that this approach exactly reproduces
the Gaussian approximation of Ref.~\cite{Pi1}, without losing the
merits of the LvN approach.

We start with the Hamiltonian:
\begin{eqnarray}
\hat{H}(t)= \frac{\hat{p}^2(t)}{2 m(t)} + V(\hat{q}(t),t),
\end{eqnarray}
where the potential is an arbitrary function of $\hat{q}(t)$ and
$t$.

Now let us assume the time-evolution of an initial ground state
can be described by a Gaussian wave-packet $\displaystyle
\Psi(q,t) = r e^{-\frac{iv^* }{2u^*}q^2-i\frac{x^*}{u^*}q  }$,
which is annihilated by an operator:
\begin{eqnarray} \label{A:def}
\hat{A} = i\left[ u^*(t) \hat{p}(t)+ v^*(t) \hat{q}(t) +
x^*\right].
\end{eqnarray}
The LvN approach is to demand this operator to be invariant under
time evolution:
\begin{eqnarray} \label{lvn}
0=\frac{d\hat{A}}{dt}=i
\left[\dot{u}^*+\frac{v^*}{m(t)}\right]\hat{p}(t) +
i\left[\dot{v}^*\hat{q}(t)-u^* V'(\hat{q},t) + \dot{x}^*\right].
\end{eqnarray}
Taking the expectation values with respect to the wave-function
$\Psi$ on each side of the equations obtained from the variations
in $\hat{p}$ or $\hat{q}$ of Eq.~(\ref{lvn}), gives
\begin{eqnarray} \label{eom}
v^*(t) = -m (t) \dot{u}(t), ~~ \dot{v}(t)- u(t) \langle
V''(\hat{q},t) \rangle =0, \\
\dot{x}^*(t) +u^*(t) \left[ \langle\hat{q}(t)\rangle \langle
V''(\hat{q},t)\rangle - \langle V'(\hat{q},t) \rangle \right]=0.
\label{dx:0}
\end{eqnarray}
These equations reduce to one equation for $u$:
\begin{eqnarray} \label{u}
m(t)\ddot{\bar{u}}+ \langle V''(\hat{q},t)\rangle \bar{u}(t) =0,
\end{eqnarray}
where $\bar{u}(t) = \sqrt{m(t)} u(t)$. This equation is too
complexly coupled to $x(t)$ to be solved exactly, except for a
simple problem of harmonic oscillators~\cite{kim2}. To overcome
this difficulty and to find a way for easier application to
physical problems, we define an auxiliary function
\begin{eqnarray} \label{barq}
\bar{q}(t) = -i \hbar \left[\bar{u}(t) x^*(t) - \bar{u}^*(t)
x(t)\right],
\end{eqnarray}
which satisfies the equation of motion
\begin{eqnarray} \label{ddq:0}
\ddot{\bar{q}}(t) + \frac{\langle
V'(\hat{q},t)\rangle}{\sqrt{m(t)}} =0.
\end{eqnarray}
The equation~(\ref{ddq:0}) becomes the ordinary differential
equation for $\bar{q}$ only in the semi-classical approximation.
In general, the expectation value $\langle F(\hat{q},t) \rangle$
depends on $\bar{q}$ and $r^2=\hbar |\bar{u}|^2$. Therefore, we
need also to consider the equation of motion for $r$~\cite{kim2}:
\begin{eqnarray} \label{ddr:0}
\ddot{r}- \frac{L^2}{r^3}+ \frac{\langle V''(\hat{q},t)
\rangle}{m(t)}r=0  ,
\end{eqnarray}
where $L=r^2(t=t_0) \dot{\phi}(t=t_0)$ is a constant of motion,
with $\phi$ being the phase part of $u(t)$. If the initial
wave-packet at origin is a static Gaussian, the initial condition
becomes $\bar{u}_i(t)=\frac{1}{\sqrt{2 \hbar
\omega_0}}e^{-i\omega_0 t}$ and $x(0)=0$, which gives $L= -1/2$.
The two equations (\ref{ddq:0}), (\ref{ddr:0}) and the boundary
condition mentioned above determine a unique solution for
$\bar{q}$ and $r$.

Once we get $u$ and $x$ from Eqs.~(\ref{eom}), (\ref{dx:0}),
(\ref{ddq:0}), and (\ref{ddr:0}), the time dependence of quantum
dynamical operators is explicitly written as:
\begin{eqnarray} \label{pq:t}
\hat{q}(t)&=& u(t) \hat{A} + u^*(t) \hat{A}^\dagger
+ \frac{\bar{q}(t)}{\sqrt{m(t)}},  \label{q:A}  \\
\hat{p}(t)&=& m(t)\left\{\dot{u}(t)\hat{A} +
\dot{u}^*(t)\hat{A}^\dagger +
\frac{d}{dt}\frac{\bar{q}(t)}{\sqrt{m(t)}}\right\}. \label{p:A}
\nonumber
\end{eqnarray}
Therefore, the expectation values of operators at time $t$ for the
state $\Psi$  are given by
\begin{eqnarray} \label{(pq):t}
\langle \hat{q}(t) \rangle &=& \frac{\bar{q}(t)}{\sqrt{m(t)}}, ~~
\langle \hat{p}(t) \rangle = m(t)\frac{d}{dt}
\frac{\bar{q}(t)}{\sqrt{m(t)}}\\
\langle \hat{q}^2(t) \rangle &=&
\left[\frac{\bar{q}(t)}{\sqrt{m(t)}}\right]^2 + \hbar^2 |u|^2, ~~
\langle \hat{p}^2(t) \rangle = \left[m(t)\frac{d}{dt}
\frac{\bar{q}(t)}{\sqrt{m(t)}}\right]^2+ m^2(t) \hbar^2
|\dot{u}|^2 .
\end{eqnarray}
To have explicit form of Eqs.~(\ref{eom}) and (\ref{ddq:0}), we
need to expand the expectation value of operator $F(\hat{q},t)$ in
series around $\bar{q}$:
\begin{eqnarray} \label{F:barq}
\langle F(\hat{q},t) \rangle = \sum_{n=0}^\infty
\frac{F^{(n)}(\bar{q}/\sqrt{m})}{n!} \langle
\left[\hat{q}(t)-\bar{q}/\sqrt{m(t)}\right]^n \rangle ,
\end{eqnarray}
where the expectation values of the powers of
$\left[\hat{q}(t)-\bar{q}/\sqrt{m(t)}\right]$ can be obtained from
the iteration formula:
\begin{eqnarray}\label{vac:iter}
\langle\left[\hat{q}(t)-\bar{q}/\sqrt{m(t)}\right]^n \rangle=
(n-1) \hbar^2 |u(t)|^2\langle\left[\hat{q}(t)
   -\bar{q}/\sqrt{m(t)}\right]^{n-2}
       \rangle ,  ~~ n \geq 2 .
\end{eqnarray}
Note that this series is the same as the series expansion in
$\hbar$ because $\hbar^2|u|^2 $ is $O(\hbar)$.

The equations ~(\ref{dx:0}) and (\ref{u}) become to $O(\hbar)$,
\begin{eqnarray}
\dot{x}^*(t) +u^*(t) \left[ \frac{\bar{q}}{\sqrt{m}} ~V^{(2)} -
V^{(1)}+ \frac{\hbar^2|\bar{u}|^2}{2m}
\left(\frac{\bar{q}}{\sqrt{m}}~ V^{(4)}- V^{(3)}\right) \right]&=&
0 , \label{dx:O1} \\
m(t)\ddot{\bar{u}}(t)+\left[V^{(2)} +
\frac{\hbar^2|\bar{u}|^2}{2m} V^{(4)}\right]
 \bar{u}(t) &=& 0 ,  \label{ddu:O1}
\end{eqnarray}
where $V^{(n)}=\partial_{x}^n V(x,t)|_{x=\bar{q}/\sqrt{m}}$ is
independent of $r$. The induced set of equations~(\ref{ddq:0}) and
(\ref{ddr:0}) is given by
\begin{eqnarray}
\ddot{r} -\frac{1}{4 r^3}+ V^{(2)} r + \frac{\hbar r^3}{2}
V^{(4)}&=& 0   \label{ddr:O1}  ,\\
 \ddot{\bar{q}}(t) +
\frac{V^{(1)}}{\sqrt{m(t)}}+\frac{\hbar r^2}{2\sqrt{m(t)}}
 V^{(3)} &=& 0  \label{ddq:O1}.
\end{eqnarray}
These equations of motion admit to define the effective
Hamiltonian,
\begin{eqnarray} \label{Heff:1}
H_{eff}(\bar{q},\bar{r}) &=&\frac{\dot{\bar{q}}^2}{2} +
\frac{\dot{\bar{r}}^2}{2} +V^{(0)} + \frac{\bar{r}^2}{2}V^{(2)} +
\frac{\hbar^2}{8 \bar{r}^2}+ \frac{\bar{r}^4 V^{(4)}}{8}  \\
  &=&\frac{\dot{\bar{q}}^2}{2} +
\frac{\dot{\bar{r}}^2}{2} +\langle V(\hat{q},t)\rangle
+\frac{\hbar^2}{8 \bar{r}^2}, \nonumber
\end{eqnarray}
where $\bar{r}= \sqrt{\hbar}r= \hbar|\bar{u}|$.
Eq.~(\ref{Heff:1}), which is a classical Hamiltonian for an
anharmonic oscillator in two dimensional space
$x^i=(\bar{q},\bar{r})$, represents the same effective potential
of Cooper, Pi, and Stancioff~\cite{Pi1}. In this sense, the
present approach exactly reproduces the variational gaussian
approximation of previous authors. The LvN method, however,
presents better understanding of the quantum mechanical system,
because not only it naturally defines the complete Fock space
through $A$ and $A^\dagger$ but also these operators are time
independent. In this sense, the parameterizations of the
wave-function by $u$ and $x$ is more convenient than by $q$ and
$r$.

From the vacuum defined by $A|0\rangle =0$, we can naturally
define the excited states by successively applying the creation
operator $A^\dagger$ on the ground state. The mixed states can
also be defined. For example, the thermal states can be defined
from the density matrix~\cite{spkim}
\begin{eqnarray} \label{rho_T}
\hat{\rho}_T(t) = \frac{1}{Z_N} e^{-\beta \hbar
\omega_0[\hat{N}(t)+ 1/2]} ,
\end{eqnarray}
where $N(t) =\hat{A}^\dagger \hat{A}  $, $\beta$ is the inverse
temperature, $\omega_0$ is the frequency of $u$ at $t=0$, and the
partition function $Z_N$ is given by
\begin{eqnarray} \label{Z_N}
Z_N = \sum_{n=0}^{\infty} \langle n|e^{-\beta \hbar
\omega_0[\hat{N}(t)+ 1/2]}|n \rangle = \frac{1}{2
\sinh\left(\frac{\beta \hbar \omega_0}{2}\right)} .
\end{eqnarray}
We explicitly write down the expectation values of some operators
with respect to the thermal state:
\begin{eqnarray} \label{op:exp}
\langle \hat{q}(t) \rangle_T &=& \bar{q}(t), ~~ \langle
\delta\hat{q}^2(t) \rangle_T = \hbar^2 |u(t)|^2 \coth
\frac{\beta \hbar \omega_0}{2} , \\
\langle \hat{p}(t) \rangle_T &=& m\dot{\bar{q}}(t), ~~\langle
\delta\hat{p}^2(t) \rangle_T = \hbar^2 m^2 |\dot{u}(t)|^2 \coth
\frac{\beta \hbar \omega_0}{2}.  \nonumber
\end{eqnarray}

\section{Slow rollover transition}
We set $m(t)=1$ in this section, since the mass dependence can be
removed by reparametrization of other parameters in the
Hamiltonian. We thus consider the potential
\begin{eqnarray} \label{V:anho}
V(\hat{q},t)= \frac{\lambda}{4!}
\hat{q}^2(t)\left\{[\hat{q}(t)-a(t)]^2+ k(t) \right\} ,
\end{eqnarray}
where we assume $a^2(t)~[ \geq -k(t)]$ increases from zero to some
constant number $a^2$, and $k(t)$ decreases from some positive
number to a fixed non-positive number $\kappa$ at $t\rightarrow
\infty$, which assures the potential at $q=0$ to be locally stable
false vacuum. The dynamics for this potential contains many
interesting quantum mechanical features depending on the
time-dependence of $k(t)$ and $a(t)$, such as tunnelling
($-a^2<\kappa <0$), phase transition ($\kappa =0$), or slow
rollover ($ \kappa = -a^2$).

The energy for a static Gaussian wave-packet at $t=0$ described
earlier is given by
\begin{eqnarray} \label{energy}
E= \frac{1}{4} \hbar \omega_0+ \frac{\lambda}{48}\hbar (a^2+k) +
\frac{\lambda \hbar^2}{32 \omega_0^2} .
\end{eqnarray}
The effective potential corresponding to Eq.~(\ref{Heff:1})
becomes
\begin{eqnarray} \label{Veff:1}
V_{eff}(\bar{q},\bar{r}) =V(\bar{q},t)
 + \frac{\lambda \bar{r}^2}{4}
 \left[\bar{q}^2-a \bar{q} +\frac{a^2+k}{6}\right] +
\frac{\hbar^2}{8 \bar{r}^2}+ \frac{\lambda \bar{r}^4 }{8} .
\end{eqnarray}
In the case of a time-independent system, we can describe the
whole dynamics in terms of this effective potential. We present a
contour plot of the effective potential in Fig. 1.
\begin{picture}(0,300)(0,0)
\includegraphics{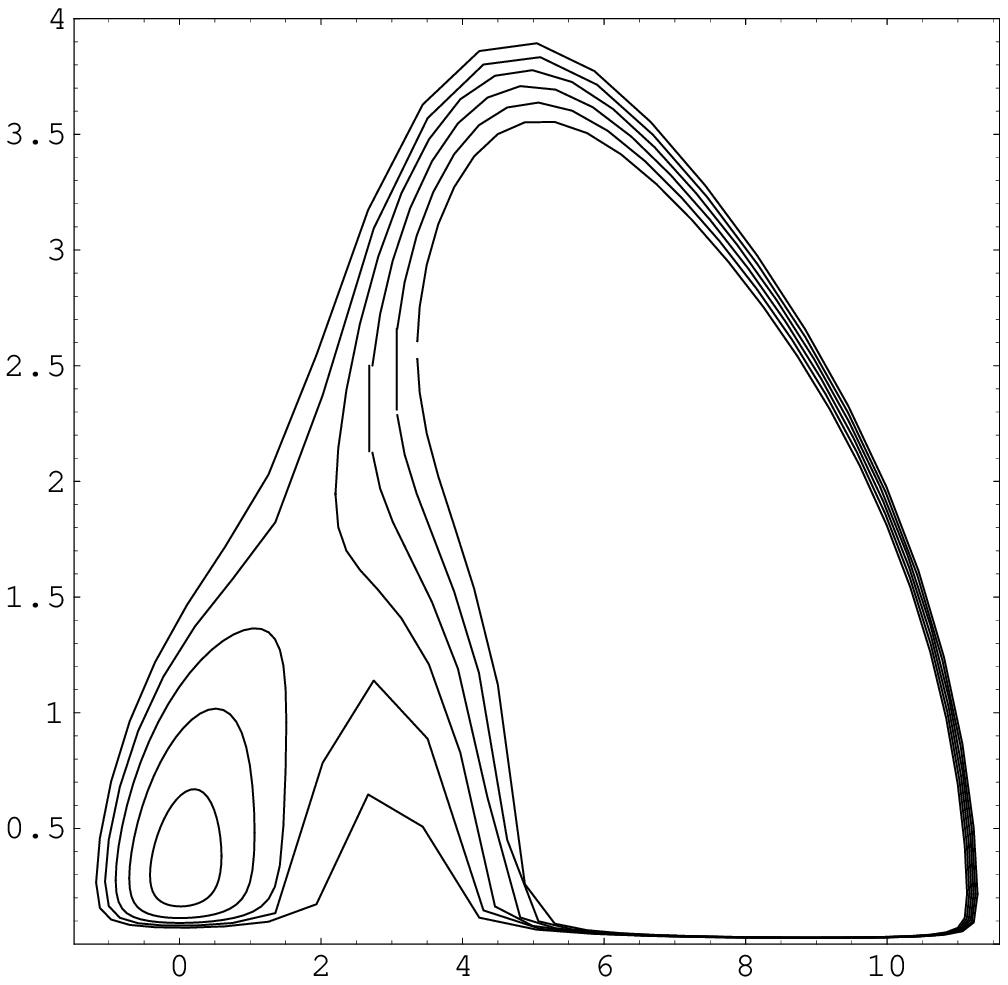}
\put(-285,260){$\bar{r}$} \put(-10,0){$\bar{q}$}
\end{picture}

{\footnotesize Fig. 1. The equipotential curves of the effective
potential. The horizontal axis represents $\bar{q}$ and the
 vertical axis $\bar{r}$. In this
figure we use the parameters, $a=8$, $k=-10$,
$\hbar^2/\lambda=0.1$, and each curve represents the energy
values, from the innermost one to outer direction, $\hbar^2
E/\lambda=$0, 1, 2, 3, 4, 5. A given Gaussian wave-function cannot
escape the contour defined by its energy.}

One crucial difference of the present paper from that of Cooper
et. al. is that we keep $\bar{r}$ dependence on the effective
potential, while it was variationally determined in
Ref.~\cite{cooper}. This results in a large difference in the slow
rolling period. As clearly shown in Fig. 1, a given wave-packet
has allowed region of $(\bar{q}, \bar{r})$ determined by the value
of its energy. Depending on the initial value of
$(\bar{q},\bar{r})$, the wave-function simply oscillates around
the false vacuum $\bar{q}=0$ or go over to the true vacuum near
$\bar{q}=8$. There are two properties which are distinguished from
its classical dynamics. First, there is a quantum mechanical
driving force at $\bar{q}=0$. The classical stable equilibrium at
$\bar{q}=0$ is not the equilibrium any more in quantum mechanics.
Second, it was shown that as the dispersion (frequency) becomes
smaller (larger), it becomes easier for $\bar{q}$ to go over the
classical potential barrier~\cite{cooper}. In Fig. 1, it is
evident that as the dispersion (frequency) becomes larger
(smaller), it also becomes easier for $\bar{q}$ to go over the
classical potential barrier. There is an inevitable restriction of
the Gaussian approximation. In the presence of leaking into the
right well the wave-function cannot be described by the Gaussian.
Due to this restriction, we cannot effectively describe tunnelling
process. To remove this restriction, one needs double Gaussian
(Appendix B of Ref.~\cite{Pi1}) or other approximations, which is
out of scope of the present paper. If $a(t)$ becomes
time-dependent, the dynamics cannot be understood simply by the
effective potential, and the equations of motion should be
considered explicitly.

As an explicit dynamical example, we consider the slow rolling
($k+ a^2=0$ for $t>0$) potential. The equipotential curves are
shown in Fig. 2. An interesting features of this effective
potential is that the unstable equilibrium point at $\bar{q}=0$
disappears.

To understand the dynamics of this system, we write down the
equation of motion for $x^i=(\bar{q}, \bar{r})$ explicitly:
\begin{eqnarray}\label{ddx:2}
\ddot{x}^i+ \nabla_i V_{eff}(x^j)=0 ,
\end{eqnarray}
where $V_{eff}(x^j)$ is given by Eq.~(\ref{Veff:1}). Let a
Gaussian wave-function prepared at $\bar{q}=0$ with velocity
$\dot{q}(0)=0$ starts to evolve. In the case of slow rollover, the
evolution is divided into three phases; the slow-rolling, rolling,
and oscillating phases.

As soon as the packet released, it starts to roll down the
potential wall. In the region $\bar{q}\sim 0$,  $\bar{r}$
increases slowly until $\bar{r} \sim [\hbar^2/(2\lambda)]^{1/6}$,
which is the variational minimum of the potential in $\bar{r}$ at
$\bar{q}=0$, because $\ddot{\bar{r}}\sim \frac{\sqrt{\hbar}}{4}
\left( 2\omega_0\right)^{3/2}$ and $\dot{\bar{r}}(0)=0$. This time
($t_1$), which $\bar{r}(t)$ takes in reaching to its variational
value, is roughly given by $t_1 \sim \frac{1}{(2 \lambda
\hbar)^{1/12}}\left(\frac{2}{\omega_0}\right)^{3/4} $. After that,
the time evolution of $(\bar{q},\bar{r})$ is similar to its
variational path. The initial time evolution of $\bar{q}$ is
governed by the quantum-mechanical driving force given by $a(t)$
until $\bar{q}^2 \simeq \bar{r}^2$. Therefore the time ($t_2$) for
$\bar{q}$ to increase to $\bar{r}$, can be obtained from
\begin{eqnarray} \label{a:sr}
\bar{q}(t_2) \simeq \frac{\lambda}{4}\int^{t_2} dt' \int^{t'}dt''
\bar{r}^2(t'') a(t'') \sim \bar{r}(t_2) \sim \bar{r}(\bar{q}).
\end{eqnarray}
\begin{picture}(200,300)(0,0)
\includegraphics{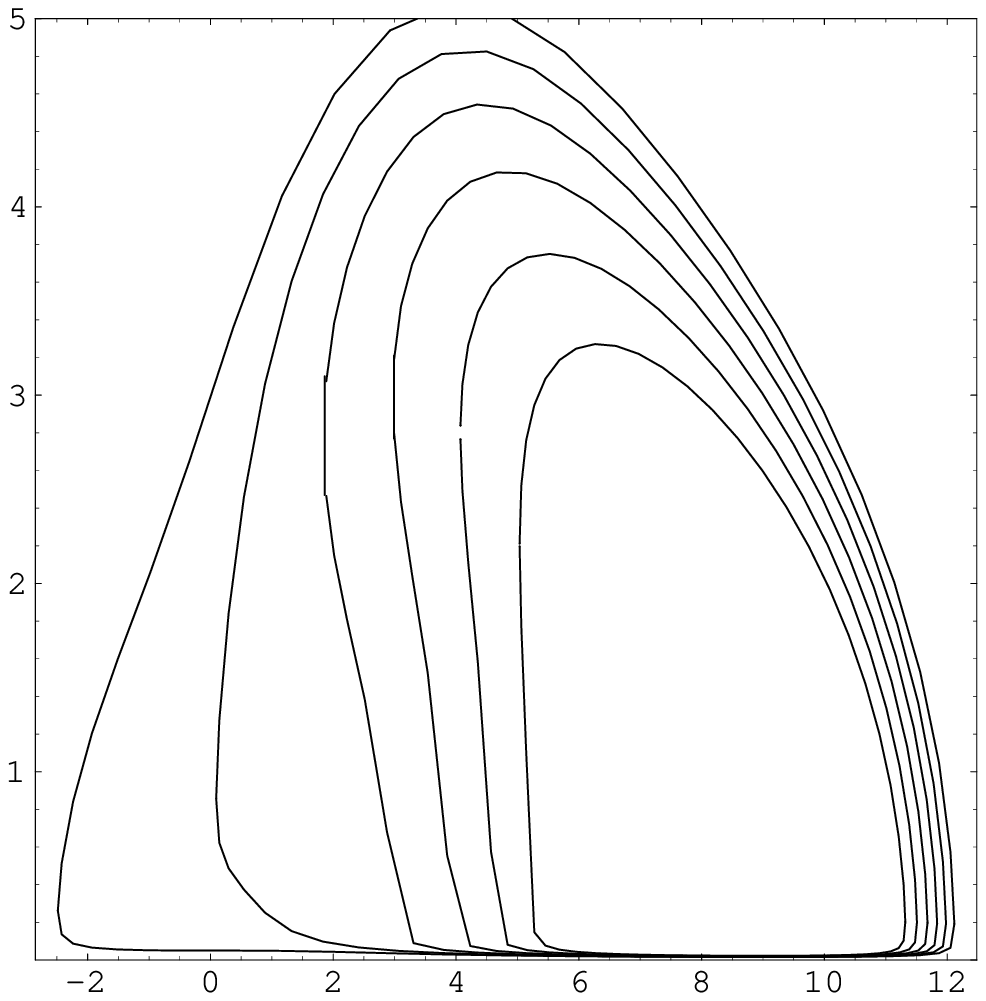}
\put(-285,260){$\bar{r}$} \put(-30,0){$\bar{q}$}
\end{picture}
\begin{flushleft}
{\footnotesize Fig. 2. The equipotential lines of a
 slow rolling transition. The
horizontal axis represents $\bar{q}$ and the
 vertical axis $\bar{r}$. In this
figure, $a=6$, $\hbar^2/\lambda=0.1$, and each curve represents
the energy values  $\hbar^2 E/\lambda=$10, 0, -10, -20, -30, -40
from outside. A given Gaussian wave-function cannot escape the
contour defined by its energy. }
\end{flushleft}

If $t_1 > t_2$, we determine $t_2$ from Eq.~(\ref{a:sr}), which
gives $t_2 \sim \left( \frac{8 \cdot 5!^2}{\lambda^2 a^2
\omega_0^3\hbar} \right)^{1/8} $. Otherwise, $t_2$ is given by the
sum of the time taken by $\bar{r}(t)$ to increase to the
variational value, $t_1$, and the variational time for the
remaining part of $\bar{q}$, $t_v$, which will be shown shortly.
The variational effective potential is $ V_{var} =
V_{eff}(\bar{q},\bar{r}(\bar{q})) $, where $\bar{r}(\bar{q})$
satisfies
\begin{eqnarray} \label{barq:barr}
\bar{r}^6(\bar{q})+ (\bar{q}^2- a
\bar{q}^2)\bar{r}^4(\bar{q})-\frac{\hbar^2}{2 \lambda} =0.
\end{eqnarray}
Therefore, in the variational approach, $\ddot{\bar{q}}(0)=
\frac{\lambda a}{4} \left(\frac{\hbar^2}{2 \lambda}\right)^{1/3},$
and the duration of slow-rolling phase is $t_v \sim
\sqrt{\frac{8}{\lambda a}}$. Note that the duration ($t_s=t_1$ or
$t_s=t_2+ t_v$) of the slow-rolling phase in both of the scenario
is considerably larger than the time determined by the variational
method $t_v$.

For $t > t_s$, the wave-packet runs into the rolling phase. In
this phase, the equation of motion becomes too difficult to get
exact answer without numerical analysis. In this paper we simply
outline the evolution without giving explicit solution. If
$a(t)=a$ is a constant for $t>t_s$, and $\bar{r}^2$ is smaller
than $\bar{q}^2$, one gets $q(t)$ from the elliptic integral
\begin{eqnarray} \label{q:t}
\int^{\bar{q}} \frac{dq}{\sqrt{c+ q^2(-2 a q + q^2)}}=
\frac{\lambda}{12} (t-t_s),
\end{eqnarray}
where $c = 12 \dot{\bar{q}}^2(t_s)/\lambda$. However, as one can
see from Eq.~(\ref{ddx:2}), $\bar{r}$ grows quickly during the
period $\bar{r}^2<a \bar{q}-\bar{q}^2$, which lasts until $\bar{q}
\simeq a$. Once $\bar{r}$ or $\bar{q}$ grows enough and fails to
satisfies the condition, $\bar{r}^2<a \bar{q}-\bar{q}^2$,
$\bar{r}$ goes into the oscillating phase.

After $\bar{q}$ reaches to the potential minima, it starts to
oscillate around the minimum of the potential, which is the last
stage. We do not restrict to the slow rolling case  in the
calculation of this stage. We simply assume that the final
wave-packet resides around the true vacuum, which is determined by
$\nabla V_{eff}|_{\vec{x}_0}=0$. For a slow rolling potential,
there is a unique point which satisfies this condition. Now, we
series expand the potential around the true vacuum:
\begin{eqnarray} \label{Veff:2}
V_{T}(\vec{x}) &=& V_{eff}(\vec{x}_0)+
\frac{1}{2}\sum_{i,j}\partial_i\partial_j
V_{eff}(\vec{x})|_{x=x_0} (x^i-x^i_{0})(x^j-x^j_{0}) +\cdots .
\end{eqnarray}
The solutions for the equation of motion for this $V_{T}$ are
given by two frequency series expansion in general:
\begin{eqnarray} \label{qr:t}
x^i(t)&=& x_{0}^i[1+ \sum_{n=1}^\infty\sum_{m=\pm}\chi_{n,m}
   \cos n(\omega_m t + \theta_m)],
\end{eqnarray}
where $\chi_{n,m}$ is determined from the differential
equation~(\ref{ddx:2}) up to its scale, and $\omega_\pm$ is the
normal mode frequencies of $V_T$. We show this fact to the first
order in the series expansion. Using the coordinates
transformation:
\begin{eqnarray} \label{x:X}
X_+ = \frac{\bar{q}+ b\bar{r}}{\sqrt{1+b^2}},~~ X_- =
\frac{b\bar{q}-\bar{r}}{\sqrt{1+b^2}},
\end{eqnarray}
the Hamiltonian, up to quadratic order, becomes
\begin{eqnarray} \label{Heff:T}
H_T = \dot{X}_+^2+ \dot{X}_-^2 + \frac{1}{2}\left[ \omega_+^2
X_+^2 + \omega_-^2 X_-^2 \right],
\end{eqnarray}
where $b$ and $\omega_\pm$ satisfy
\begin{eqnarray} \label{bom}
&&b^2- \frac{\partial_q^2 V_{eff}(\vec{x})|_{x=x_0}-\partial_r^2
V_{eff}(\vec{x})|_{x=x_0} }{2\partial_1\partial_2
V_{eff}(\vec{x})|_{x=x_0}  }b-1=0 ,  \\
&& \omega^2_\pm = \frac{\sum_i
\partial_i^2 V_{eff}(\vec{x})|_{x=x_0}}{2}-\frac{1+b^2}{b}
\partial_1\partial_2 V_{eff}(\vec{x})|_{x=x_0} .
\end{eqnarray}
The solutions for the Hamiltonian system~(\ref{Heff:T}) are $
X_\pm(t) = X_{0,\pm } \cos(\omega_\pm t + \theta_\pm)$, where
$X_{0,\pm}$ should be determined from the whole time evolution.
From this we get, to first order in $X_0$,
\begin{eqnarray} \label{qr:O1}
\bar{q}(t) = q_0+\frac{X_+(t)+ bX_-(t)}{\sqrt{1+ b^2}},
~~\bar{r}(t) = r_0+\frac{bX_+(t)- X_-(t)}{\sqrt{1+ b^2}} .
\end{eqnarray}
We have two unknown parameters $X_{0,\pm}$ which should be
determined from the whole time-evolution. If the system is time
independent for $t>0$, one of them can be determined from the
energy conservation law.

We now turn our attention to the calculation of $\bar{u}(t)=r
e^{-i \phi}/\sqrt{\hbar}$. This can be done by integrating
$\dot{\phi} = 1/(2 r^2)$. But, this integration is highly
nontrivial if $X_0$ is not much smaller than $r_0$, and the slow
rolling is the case. As a round about method, we directly try to
solve the equation for $u$, and then determine the free parameters
in $u$ from the effective potential. To simplify the procedures,
we assume $b \ll 1$ and $r_0 \ll q_0$. These two assumptions are
usually satisfied for large $\bar{q}_0$. Now we write
\begin{eqnarray} \label{barq:1}
\bar{q}=q_0[1-\chi \cos(\omega t+\theta)],
\end{eqnarray}
where $q_0 \chi=-X_{0,+}$. The equation~(\ref{ddu:O1}) for $u$, to
linear order in $\chi$, can be approximated to be:
\begin{eqnarray} \label{eom:v2}
\ddot{\bar{u}}(t) + \omega^2[1-\chi'  \cos(\omega
t+\theta)]\bar{u}(t) =0,
\end{eqnarray}
where $\displaystyle \chi'= \frac{\lambda (q_0^2-a
q_0/2)}{\omega^2}\chi $. With the transformation $\omega t+ \theta
= 2 \phi$, Eq.~(\ref{eom:v2}) becomes the Mathieu equation with
$s=2, ~~ b= 4+4 \chi',$ and $h^2= 8\chi'$ in page 562 of
Ref.~\cite{morse}. Its solution, to the first order in $\chi'$, is
given by
\begin{eqnarray}\label{v:t}
\bar{u}(t)= \alpha u_+(t) + \beta^* u_+^*(t),
\end{eqnarray}
where $\alpha$ and $\beta$ are constants, and $\displaystyle
u_+(t) = \frac{1}{\sqrt{2\hbar \omega}}\left[ e^{-i (\omega
t+\theta)}- \frac{\chi'}{6}e^{-2i (\omega t + \theta)}+
\frac{\chi'}{2} \right]$. Note that $u_+$ satisfies $u_+
\dot{u}_+^*- u_+^* \dot{u}_+ = i$. Note also that
\begin{eqnarray} \label{u2:t}
\bar{r}^2(t)= \hbar^2 |u(t)|^2 = \hbar^2[(|\alpha|^2+ |\beta|^2)
|u_+(t)|^2 + \alpha \beta u_+^2 +\alpha^* \beta^* u_+^{*2}] ,
\end{eqnarray}
where its time average $\hbar(|\alpha|^2+ |\beta|^2)/(2\omega)$
should give $\bar{r}_0^2$. From this relation and
Eq.~(\ref{x:al:be}) below, we determine $|\beta|= \sqrt{\omega
r_0^2/\hbar -1/2}$. Yet undetermined parameters are $\chi$, the
size of fluctuation of $\bar{q}$, and the relative phase between
$\alpha$ and $\beta$, whose determinations need the analysis of
whole time evolution.

After integrating the equation~(\ref{dx:0}) in $t$  one obtains
\begin{eqnarray}\label{x:t}
x(t) = x_0+q_0 \left[\alpha \dot{u}_+(t) +\beta^*
\dot{u}_+^*(t)\right] .
\end{eqnarray}
By equating $-i\hbar[ u(t) x^*(t)-u^*(t) x(t)]$ computed by (39)
and (41) to $\bar{q}(t)$ in~(\ref{barq:1}), one obtains the
following two equations:
\begin{eqnarray}\label{x:al:be}
|\alpha|^2 - |\beta|^2 &=& 1, \\
x_0&=&-i \sqrt{\frac{ \omega}{2\hbar}} q_0\chi (\alpha-\beta^*),
\nonumber
\end{eqnarray}
where $x_0$ denotes the constant shifting of annihilation
operator, which gives a constant coherence, and $\beta$ denotes
the squeezing with respect to the true vacuum.

The invariant annihilation operator is related to the annihilation
and creation operators at the true vacuum by the following
relation:
\begin{eqnarray} \label{A:a}
\hat{A}&=& i \left[\bar{u}^* \hat{p}- \dot{\bar{u}}^*
\hat{q} + x^* \right] \\
 &=& \alpha^* \hat{a} + \beta \hat{a}^\dagger  , \nonumber
\end{eqnarray}
where $ \hat{a} = i\left[ u_+^*(t) \hat{p}(t)-\dot{u}_+^*(t)
(\hat{q}(t) -q_0)+ \sqrt{\frac{\omega}{2 \hbar}} q_0 \chi\right]$
is the annihilation operator at $\bar{q}=q_0$ with $O(\chi^1)$
constant coherence. We cannot determine all the parameters without
solving the full time evolution of the system, which needs
numerical work. However, many important parts of the dynamics are
determined without explicitly solving the full dynamics, e.g., the
squeezing and coherence are determined just by the effective
potential at $t \rightarrow \infty$, which is new results.

\section{Summary and discussions}
We have developed a LvN approach which gives the time-dependent
Gaussian approximation. We, then, applied the method to a system
with time-dependent slow rolling type potential, and found the
time evolution of a Gaussian wave-packet initially centered at
$q=0$. We divided the evolution into three stages, and described
solutions for each stages. We have calculated the period of the
slow rolling phase ($t_s$), which is significantly larger compared
to the variationally determined value $t_v= \sqrt{8/(\lambda a)}$.
We also displayed that many information of the final stage can be
obtained without explicit solution of the full time evolution.
Explicitly, we have shown that the invariant annihilation operator
can be expressed as a linear combination of the creation and
annihilation operators of true vacuum. We have also shown that the
squeezing and constant coherence are determined only by the
dynamics around the absolute minimum of the effective potential,
independent of the details of the whole dynamics.

This approach is the synthesis of the merits of the Gaussian
approximation and LvN method. In the point of view of Gaussian
approximation, it shows exactly the same potential in $\bar{q}$
and $\bar{r}$ as that of Ref.~\cite{Pi1}, and in the point of view
of LvN method, it was shown that the method can be applicable for
general type of potential as a well posed approximation. There is,
however, restrictions for the application range of the type of the
potential. The Gaussian approach is not appropriate to describe
the symmetry breaking potential from the start to the end, even
though it is fairly accurate until $\bar{r} \simeq
\sqrt{\frac{2}{3}}a$. This limitation shown in Ref.~\cite{Pi1} was
not cured in this paper. To overcome this limitation we need
double Gaussian trial wave function or other approximations.

\vspace{0.5cm}
~\\
{\Large {\bf Acknowledgments}} \\
~\\
This work was supported in part by Korea Research Foundation under
Project number KRF-2001-005-D00010 (H.-C.K. and J.H.Y.).

\vspace{0.5cm}
~\\
\newpage


\end{document}